\documentstyle[12pt]{article}
\renewcommand{\theequation}{\thesection.\arabic{equation}}

\begin{document}
\begin{titlepage}
\vskip 2.00cm
\title{\bf Motion of Colored Particle in a Chromomagnetic Field}
\author{Sh.Mamedov\thanks{Email: sh$_-$mamedov@yahoo.com}\\
{\small {\em High Energy Physics Lab., Baku State University, }}\\
{\small {\em Z.Khalilov str.23, Baku 370148, Azerbaijan}}\\
{\small {\em and }} \\
{\small {\em Institute for Studies in Theoretical Physics and
Mathematics (IPM),}} \\
{\small {\em P.O.Box 19395-5531, Tehran, Iran}}}
\date{}
\maketitle
\begin{abstract}
The Dirac equation in a chromomagnetic field is solved for colored particle
moving in a limited space volume. Quantized energy levels and the corresponding
wave functions are found for backgrounds both directed along third axes
and having spherical symmetry. It was shown interrelation with the case of
motion in an infinite space volume.
\end{abstract}
\end{titlepage}
\hspace{20mm}
\newline
{\Large {\bf Introduction}}
\vspace{4mm}
\newline
Quantum and classical mechanics of non-abelian charged particles has been
studied by various authors. Problems in this theory serve to find the solution of QCD
problems and are connected with description of quark's motion inside hadrons
or any other limited space volume. Since QCD has confinement property,
presumably it will be better to consider QCD problems in a finite volume as
well [1]. As is known, there are color background fields in the QCD vacuum [2]
and so, it makes sense to study the problem of motion of a colored particle in
a color background in limited space volume. The simplest case, a constant
color background could be given by two different types of vector potentials
[3,4]. Problems in an external color field given by abelian vector potentials
are solved analoguosly to the ones in abelian theory. The second type of
vector potentials are non-commuting constant vector potentials, which are
not gauge equivalent to first type [4]. These potentials are used
to solve the ground state problem of QCD [4,5] and other different problems
connected with the QCD vacuum [5-9] as well and sometimes give physically
different results from that ones in which first type
potensials was used. Here we shall solve Dirac equation in a finite space volume in a
constant background chromomagnetic field given by the second type vector potentials. The
solutions and spectra found are more suited to the physical situation and by use of them 
could be constructed quark's Green's function by the help of exact solution
method [10], which could be used in solving QCD problems.

\section{ Constant homogenous background}
\setcounter{equation}{0}
Let us define an external chromomagnetic field by constant vector potentials.
Within the SU(3) color symmetry group they look like

\begin{equation}
\label{1.1}A_1^a=\sqrt{\tau }\delta _{1a},\ A_2^a=\sqrt{\tau }\delta _{2a},\
A_3^a=0,\ A_0^a=0, 
\end{equation}
where $a=\overline{1,8}$ is a color index, $\tau $ is a constant and $\delta
_{\mu a}$ is the Kroneker symbol. The field (1.1) is directed along the third axes
of ordinary and color spaces :

\begin{equation}
\label{1.2}F_{12}^3=H_z^3=g\tau \quad other\ \ F_{\mu \nu }^a=0. 
\end{equation}
Here $g$ is the color interaction constant.

The Dirac equation for a colored particle in the external color field has the
form

\begin{equation}
\label{1.3}\left( \gamma ^\mu P_\mu -M\right) \psi =0, 
\end{equation}
where $P_\mu =p_\mu +gA_\mu =p_\mu +gA_\mu ^a\frac{\lambda ^a}2$ , $\lambda
^a$ are Gell-Mann matrices describing particle's color spin. Equation
(1.3) written for Maiorana spinors $\phi $ and $\chi $ has the well-known form 
\begin{equation}
\label{1.4}\left( \sigma ^iP_i\right) ^2\psi =-\left( \frac{\partial ^2}{%
\partial t^2}+M^2\right) \psi , 
\end{equation}
where the Pauli matrices $\sigma ^i$ describe a particle's spin. Here and afterwards $%
\psi $ means $\phi $ or $\chi .$ The spinors $\phi $ and $\chi $ have two
components, corresponding to the two spin states of a particle $\psi =\left( 
\begin{array}{c}
\psi _{+} \\ 
\psi _{-} 
\end{array}
\right) .$ Each component of $\psi $ transforms under the fundamental
representation of color group SU(3) and has three color components
describing color states of a particle corresponding to three eigenvalues
of color spin $\lambda ^3$

\begin{equation}
\label{1.5}\psi _{\pm }=\left( 
\begin{array}{c}
\psi _{\pm }(\lambda ^3=+1) \\ 
\psi _{\pm }(\lambda ^3=-1) \\ 
\psi _{\pm }(\lambda ^3=0) 
\end{array}
\right) =\left( 
\begin{array}{c}
\psi _{\pm }^{(1)} \\ 
\psi _{\pm }^{(2)} \\ 
\psi _{\pm }^{(3)} 
\end{array}
\right) . 
\end{equation}

Since the background field (1.1) is time independent, the equation (1.4) has got
the form 
\begin{equation}
\label{1.6}\left[ \overrightarrow{p}^2+\frac{g^2\tau }2I_2+g\tau ^{\frac
12}\left( p_1\lambda ^1+p_2\lambda ^2\mp \frac{H_z^3}2\lambda ^3\right)
\right] \psi _{\pm }=\left( E^2-M^2\right) \psi _{\pm }. 
\end{equation}
Here $I_2=\left( 
\begin{array}{ccc}
1 & 0 & 0 \\ 
0 & 1 & 0 \\ 
0 & 0 & 0 
\end{array}
\right) $ is the color matrix. Writing this equation for the color components (1.5)
we get two independent systems of differential equations 
\begin{equation}
\label{1.7} 
\begin{array}{c}
\left\{ 
\begin{array}{c}
\overrightarrow{p}^2\psi _{+}^{(1)}+g\tau ^{\frac 12}\left( p_1-ip_2\right)
\psi _{+}^{(2)}=\left( E^2-M^2\right) \psi _{+}^{(1)} \\ \left( 
\overrightarrow{p}^2+g^2\tau \right) \psi _{+}^{(2)}+g\tau ^{\frac 12}\left(
p_1+ip_2\right) \psi _{+}^{(1)}=\left( E^2-M^2\right) \psi _{+}^{(2)} \\ 
\overrightarrow{p}^2\psi _{+}^{(3)}=\left( E^2-M^2\right) \psi _{+}^{(3)} 
\end{array}
\right. \\ 
\left\{ 
\begin{array}{c}
\left( 
\overrightarrow{p}^2+g^2\tau \right) \psi _{-}^{(1)}+g\tau ^{\frac 12}\left(
p_1-ip_2\right) \psi _{-}^{(2)}=\left( E^2-M^2\right) \psi _{-}^{(1)} \\ 
\overrightarrow{p}^2\psi _{-}^{(2)}+g\tau ^{\frac 12}\left( p_1+ip_2\right)
\psi _{-}^{(1)}=\left( E^2-M^2\right) \psi _{-}^{(2)} \\ \overrightarrow{p}%
^2\psi _{-}^{(3)}=\left( E^2-M^2\right) \psi _{-}^{(3)} 
\end{array}
\right. 
\end{array}
\end{equation}
From (1.7) we get the same equation for all states $\psi _{\pm }^{(i)}$ $%
(i=1,2)$ 
\begin{equation}
\label{1.8}\left[ \left( \overrightarrow{p}^2+\frac{g^2\tau }%
2+M^2-E^2\right) ^2-g^2\tau \left( p_{\bot }^2+\frac{g^2\tau }4\right)
\right] \psi _{\pm }^{(i)}=0, 
\end{equation}
where $p_{\bot }^2=p_1^2+p_2^2.$ This equation possesses rotational
invariance around the $z$ axis. If we consider free motion of the particle in
all $\overrightarrow{x}$ space, then we should not impose any boundary
condition on solution of (1.8) and shall get continous energy spectra
found in [11,12] for plane wave solutions $\psi _{\pm }^{(i)}\sim e^{i 
\overrightarrow{p}\overrightarrow{r}}$ 
\begin{equation}
\label{1.9}E_{1,2}^2=\left( \sqrt{p_{\bot }^2+\frac{g^2\tau }4}\mp \frac{%
g\tau ^{\frac 12}}2\right) ^2+p_3^2+M^2. 
\end{equation}
This spectrum is used in solving of various problems [5-9]. Since free
motion of the colored particle in nature could be take place in limited
space, for example inside a hadron, let us consider motion of particle in
limited space bounded by a cylinder with radius $r_0$ and hight $z_0$ and
solve (1.8) on boundary conditions $\psi _{\pm }^{(i)}\left(
r=r_0,z\right) =0,\ \psi _{\pm }^{(i)}\left( r,z_0\right) =0$ $\left(
r^2=x^2+y^2\right) .$ Let us rewrite (1.8) in the form 
\begin{equation}
\label{1.10} 
\begin{array}{c}
\left[ \left( -\nabla ^2+ 
\frac{g^2\tau }4\right) ^2-2\left( E^2-M^2+\frac{g^2\tau }4\right) \left(
-\nabla ^2+\frac{g^2\tau }4\right) +\right. \\ \left. \left( E^2-M^2-\frac{%
g^2\tau }4\right) ^2-g^2\tau \frac{\partial ^2}{\partial z^2}\right] \psi
_{\pm }^{(i)}=0. 
\end{array}
\end{equation}
With $\left( -\nabla ^2+\frac{g^2\tau }4\right) \psi _{\pm
}^{(i)}=\eta _{\pm }^{(i)}$ it has got the following form 
\begin{equation}
\label{1.11} 
\begin{array}{c}
\left( -\nabla ^2+ 
\frac{g^2\tau }4\right) \eta _{\pm }^{(i)}-2\left( E^2-M^2+\frac{g^2\tau }%
4\right) \eta _{\pm }^{(i)}+ \\ \left( \left( E^2-M^2-\frac{g^2\tau }%
4\right) ^2-g^2\tau \frac{\partial ^2}{\partial z^2}\right) \psi _{\pm
}^{(i)}=0. 
\end{array}
\end{equation}
Acting on equation (1.11) by operator $\left( -\nabla ^2+\frac{g^2\tau }%
4\right) $ we get an equation for $\eta _{\pm }^{(i)}$ that has the same form as for 
$\psi _{\pm }^{(i)},$ i.e., (1.10) with the replacement $\psi _{\pm }^{(i)}$ on $%
\eta _{\pm }^{(i)}.$ This means that $\eta _{\pm }^{(i)}$ and $\psi _{\pm }^{(i)}$
differs only by a general constant multiplier $k^2:$ $\eta _{\pm
}^{(i)}=k^2\psi _{\pm }^{(i)}$ or 
\begin{equation}
\label{1.12}\left( -\nabla ^2+\frac{g^2\tau }4\right) \psi _{\pm
}^{(i)}=k^2\psi _{\pm }^{(i)}.
\end{equation}
Equation (1.12) is equivalent to (1.10). In the appendix is shown
another way to reduce (1.10) to the form (1.12). In a cylindrical coordinate
system the solution of (1.12) can be looked for using the separation
ansatz $\psi _{\pm }^{(i)}\left( \overrightarrow{r}\right) =\psi \left(
r\right) \cdot u\left( \varphi \right) \cdot v\left( z\right) $ and
(1.12) is divided into three independent equations 
\begin{equation}
\label{1.13}\left\{ 
\begin{array}{c}
\frac 1r\frac \partial {\partial r}\left( r 
\frac{\partial \psi \left( r\right) }{\partial r}\right) +\left( k^2-\frac{%
g^2\tau }4-\lambda ^2-\frac{m^2}{r^2}\right) \psi \left( r\right) =0 \\ - 
\frac{\partial ^2u\left( \varphi \right) }{\partial \varphi ^2}=m^2u\left(
\varphi \right) \\ \frac{\partial ^2v\left( z\right) }{\partial z^2}+\lambda
^2v\left( z\right) =0 
\end{array}
\right. 
\end{equation}
The solutions of last two equations with the boundary condition $v\left(
z_0\right) =0$ are $u\left( \varphi \right) =\frac 1{\sqrt{2\pi }
}e^{im\varphi }$ $( m=0,\pm 1,\pm 2,...$ and $v\left( z\right)
=\sin p_3z\left( \lambda ^2=p_3^2,\ p_3=\frac{\pi n}{z_0}\right) .$ For the $%
p_3=0$ case we can choose $v\left( z\right) =1.$ It is worth to remark that
 (1.12) and (1.10) become equivalent after taking the last
equation in them into account and there is no such problem for the case $%
p_3=0.$ Taking into account explicit expression of $v\left( z\right) $ and
(1.12) in (1.10) we get the following equation for the constant $k^2$%
$$
\left( k^2\right) ^2-2k^2\left( E^2-M^2+\frac{g^2\tau }4\right) +\left(
E^2-M^2-\frac{g^2\tau }4\right) ^2+g^2\tau p_3^2=0, 
$$
from which one finds the following expression for the constant $k^2$%
\begin{equation}
\label{1.14}k_{1,2}^2=\left( \sqrt{E^2-M^2-p_3^2}\pm \frac{g\tau ^{\frac 12}}%
2\right) ^2+p_3^2. 
\end{equation}
The first equation in (1.13) is the Bessel equation. Under the finitness condition on $%
r\rightarrow 0$ it has a solution determined by the Bessel function 
\begin{equation}
\label{1.15}\psi _{1,2}\left( r\right) =J_m\left( r\sqrt{k_{1,2}^2-p_3^2- 
\frac{g^2\tau }4}\right) . 
\end{equation}
If we do not impose any boundary condition on the states (1.15) with given $m,$
then they have an energy from continous spectrum (1.9), which does not depend on $%
m.$ Imposing the boundary condition $\psi _{1,2}\left( r_0\right) =0$ we find two
branchs of quantized energy levels of the colored particle in a chromomagnetic
field (1.1) 
\begin{equation}
\label{1.16}\left( E_m^{(N)}\right) _{1,2}^2=\left( \sqrt{\frac{\left(
\alpha _m^{(N)}\right) ^2}{r_0^2}+\frac{g^2\tau }4}\mp \frac{g\tau ^{\frac
12}}2\right) ^2+p_3^2+M^2 .
\end{equation}
Here $\alpha _m^{(N)}$ are the zeros of Bessel function and the quantity $N$
labels the sequence of zeros $N=1,2,3...$. From (1.16) one sees that
these energy branchs, as branches of the continous spectrum (1.9), are defined
by $\pm $ sign of the field strength, which is not connected with direction of
field. The energy levels $E_m^{(N)}$ are determined by the $m$ -chromomagnetic
quantum number$,$ i.e. by the projection of the chromomagmetic moment of the
particle onto the chromomagnetic field. Another quantum number is $N$,
determining the energy level $E_m^{(N)}$, a result of quantization due to the finitness of the motion. In contrast to the discrete spectum (1.16) the continous spectum
(1.9) is not determined by the quantum number $m.$ Thus, finitness of space
volume turns continous spectra (1.9) for a given $m$ into a discrete series
determined by the quantum number $N.$ Both spectra (1.9) and (1.16) are infinitely
degenerate. Since (1.12) is universal for the spin $\pm $ and color $\left(
i\right) $ indices, any of these states may have any energy from spectra
(1.16) or (1.9). If in the energy branchs (1.16) one makes the replacement $\frac{%
\alpha _m^{(N)}}{r_0}=p_{\bot }$ we get continous spectra (1.9). This is suitable
to quantization of momentum in standing waves. Taking into account spectrum
(1.16) in (1.14) we find the value of constant $k^2$,%
$$
k_{1,2}^2=\left( k_m^{\left( N\right) }\right) ^2=\frac{\left( \alpha
_m^{(N)}\right) ^2}{r_0^2}+\frac{g^2\tau }4+p_3^2, 
$$
which is the same for both energy branchs and so, the solution (1.15) is the same for
these branchs. Thus, the wave functions $\psi _{\pm }^{(i)}\left( 
\overrightarrow{r}\right) $ from chosen energy branch $\left(
E_m^{(N)}\right) _1$or $\left( E_m^{(N)}\right) _2$ are equal to 
\begin{equation}
\label{1.17}\psi _{\pm }^{(i)}\left( \overrightarrow{r}\right)
=\sum_{m=-\infty }^{+\infty }\frac 1{\sqrt{2\pi }}e^{im\varphi }\sin \left(
p_3z\right) J_m\left( r\frac{\alpha _m^{(N)}}{r_0}\right) . 
\end{equation}

Thus, the motion of colored particle in the field (1.1) along $z$ axis is
a standing wave and in the $\left( x,y\right) $ plane is a rotation. Writing (1.13) for the radius $a$ of the cirle in which particle rotates%
$$
\frac 1a\frac \partial {\partial r}\left( a\frac{\partial \psi \left(
a\right) }{\partial r}\right) +\left( k^2-\frac{g^2\tau }4-p_3^2-\frac{m^2}{%
a^2}\right) \psi \left( a\right) =0 
$$
we can find discrete values for these orbits%
$$
a_m^{(N)}=\frac m{\alpha _m^{(N)}}r_0. 
$$
In the continous spectrum case the constant $k^2$ and radius $a$ are 
$$
k^2=\overrightarrow{p}^2+\frac{g^2\tau }4,\ a=\frac m{p_{\bot }} 
$$
and the solution (1.15) becomes%
$$
\psi _{1,2}\left( r\right) =\psi \left( r\right) =J_m\left( rp_{\bot
}\right) . 
$$
The states $\psi _{\pm }^{(3)}\left( \overrightarrow{r}\right) $ correspond to
states of freely moving colorless scalar particle.

\section{Spherically symmetric configuration of background}
\setcounter{equation}{0}

Let us consider the case of spherical symmetry in the ordinary space with
constant chromomagmetic field defined by non-commuting potentials 
\begin{equation}
\label{2.1}A_1^a=\sqrt{\tau }\delta _{1a},\ A_2^a=\sqrt{\tau }\delta _{2a},\
A_3^a=\sqrt{\tau }\delta _{3a},\ A_0^a=0. 
\end{equation}
The non-zero components of the field strength tensor are 
$$
F_{23}^1=F_{31}^2=F_{12}^3=g\tau . 
$$
In the external field (2.1) the explicit form of (1.4) is 
\begin{equation}
\label{2.2}\left( \overrightarrow{p}^2+M^2+\frac{3g^2\tau }4+g\tau ^{\frac
12}\lambda ^ap^a-\frac{g^2\tau }2\sigma ^a\lambda ^a\right) \Psi =E^2\Psi. 
\end{equation}
Let us denote components of the spinor $\Psi $ by $\psi _1$ and $\psi _2$ : $\Psi
=\left( 
\begin{array}{c}
\psi _1 \\ 
\psi _2 
\end{array}
\right) .$ Since the external field has three components in both spaces, in this
case, different from the previous one, the components $\psi _{1,2}^{\left(
a\right) }$ do not describe states with definite value of the spin and color
spin projection onto the field. Equation (2.2) can be written in terms of a system of
differential equations for the components $\psi _{1,2}^{\left( a\right) }$%
\begin{equation}
\label{2.3}\left\{ 
\begin{array}{c}
\left( A+g\tau ^{\frac 12}p_3- 
\frac{g^2\tau }2\right) \psi _1^{(1)}+g\tau ^{\frac 12}\left(
p_1-ip_2\right) \psi _1^{(2)}=0 \\ \left( A-g\tau ^{\frac 12}p_3+ 
\frac{g^2\tau }2\right) \psi _1^{(2)}+g\tau ^{\frac 12}\left(
p_1+ip_2\right) \psi _1^{(1)}=g^2\tau \psi _2^{(1)} \\ \left( 
\overrightarrow{p}^2+M^2\right) \psi _1^{(3)}=E^2\psi _1^{(3)} \\ \left(
A+g\tau ^{\frac 12}p_3+ 
\frac{g^2\tau }2\right) \psi _2^{(1)}+g\tau ^{\frac 12}\left(
p_1-ip_2\right) \psi _2^{(2)}=g^2\tau \psi _1^{(2)} \\ \left( A-g\tau
^{\frac 12}p_3- 
\frac{g^2\tau }2\right) \psi _2^{(2)}+g\tau ^{\frac 12}\left(
p_1+ip_2\right) \psi _2^{(1)}=0 \\ \left( \overrightarrow{p}^2+M^2\right)
\psi _2^{(3)}=E^2\psi _2^{(3)} 
\end{array}
\right. , 
\end{equation}
where operator $A$ denotes $A=\overrightarrow{p}^2+M^2+\frac{3g^2\tau }%
4-E^2.$ From the system (2.3) we get the same differential equation for the all
states $\psi _{1,2}^{(i)}$ $\left( i=1,2\right) $%
\begin{equation}
\label{2.4}\left[ \left( A-\frac{g^2\tau }2\right) ^2-g^2\tau 
\overrightarrow{p}^2\right] \left[ \left( A+\frac{g^2\tau }2\right)
^2-g^2\tau \left( \overrightarrow{p}^2+g^2\tau \right) \right] \psi
_{1,2}^{(i)}=0, 
\end{equation}
which possesses rotational invariance. Since the operator in the first square
brackets commutes with the second one, (2.4) could be divided into two
equations 
\begin{equation}
\label{2.5} 
\begin{array}{c}
\left[ \left( A- 
\frac{g^2\tau }2\right) ^2-g^2\tau \overrightarrow{p}^2\right] \psi
_{1,2}^{(i)}=0, \\ \left[ \left( A+\frac{g^2\tau }2\right) ^2-g^2\tau \left( 
\overrightarrow{p}^2+g^2\tau \right) \right] \psi _{1,2}^{(i)}=0. 
\end{array}
\end{equation}
From (2.5) we get four branchs of the continous energy spectrum found in [11]
for plane wave solutions $\psi _{1,2}^{(i)}\sim e^{i\overrightarrow{p} 
\overrightarrow{r}}$ 
\begin{equation}
\label{2.6} 
\begin{array}{c}
E_{1,2}^2=\left( 
\sqrt{\overrightarrow{p}^2}\mp \frac{g\tau ^{\frac 12}}2\right) ^2+M^2, \\ 
E_{3,4}^2=\left( \sqrt{\overrightarrow{p}^2+g^2\tau }\mp \frac{g\tau ^{\frac
12}}2\right) ^2+M^2.  
\end{array}
\end{equation}
In the same manner as used in the previous section from (2.5) we get the
equivalent equations 
\begin{equation}
\label{2.7}\left( -\bigtriangledown ^2\right) \psi _j^{(i)}=k_{1,2}^2\psi
_j^{(i)},\qquad \left( -\bigtriangledown ^2+g^2\tau \right) \psi
_j^{(i)}=k_{3,4}^2\psi _j^{(i)}\qquad \left( i,j=1,2\right) . 
\end{equation}
If we take (2.7) into account corresponding equations (2.5) we
found values of $k_d^2$%
\begin{equation}
\label{2.8}k_{1,2}^2=k_{3,4}^2=k^2=\left( \sqrt{E^2-M^2}\pm \frac{g\tau
^{\frac 12}}2\right) ^2 .
\end{equation}
The solution of (2.7) can be found in textbooks of quantum mechanics (
[13] ). In a spherical coordinate system for the wave functions $\psi _j^{(i)}$
one used separation ansatz, i.e. it is chosen as a multiplication of the radial $%
R(r) $ and angular $Y_l^m\left( \theta ,\varphi \right) $ parts%
$$
\psi _j^{(i)}\left( \overrightarrow{r}\right) =R(r)\cdot Y_l^m\left( \theta
,\varphi \right) .
$$
Here $r=\sqrt{x^2+y^2+z^2},$ $l$ and $m$ are the orbital angular momentum and
chromomagnetic quantum numbers, $\theta ,\varphi $ are the polar and azimuthal
angles, respectively. The spherical functions $Y_l^m\left( \theta ,\varphi \right) $ are
expressed by means of Legendre polynomials $P_l^{\mid m\mid }\left( \cos
\theta \right) $%
\begin{equation}
\label{2.9}Y_l^m\left( \theta ,\varphi \right) =a_m\sqrt{\frac{\left(
2l+1\right) }{4\pi }\frac{\left( l-\mid m\mid \right) !}{\left( l+\mid m\mid
\right) !}}P_l^{\mid m\mid }\left( \cos \theta \right) e^{im\varphi } 
\end{equation}
with

$a_m=\left\{ 
\begin{array}{c}
1\ \ on\ m\geq 0 \\ 
\left( -1\right) ^m\ \ m<0 
\end{array}
\right. .$

The equations for the radial parts of wave functions are the same as for many quantum
mechanical problems possessing rotational invariance 
\begin{equation}
\label{2.10}\bigtriangledown _r^2R(r)+\left( k_\nu ^2-\frac{l\left(
l+1\right) }{r^2}\right) R(r)=0, 
\end{equation}
where $\nu =1,2,3,4;\ k_{1,2}^2=k^2,\ k_{3,4}^2=k^2-g^2\tau .$ With the notations $%
Q\left( r\right) =\sqrt{r}R(r)$ (2.10) turns into Bessel's
equations for $Q\left( r\right) $ 
\begin{equation}
\label{2.11}Q^{\prime \prime }\left( r\right) +\frac 1rQ^{\prime }\left(
r\right) +\left( k_\nu ^2-\frac{\left( l+\frac 12\right) ^2}{r^2}\right)
Q\left( r\right) =0. 
\end{equation}
The function $R(r)$ must be finite on $r\rightarrow 0.$ This means that solution
of (2.11) are first type Bessel functions 
\begin{equation}
\label{2.12}R_l^\nu =\frac{C_l^\nu }{k_\nu \sqrt{r}}J_{l+1/2}\left( k_\nu
r\right) . 
\end{equation}
Imposing boundary conditions $R_l^\nu (r_0)=0$ we enclose motion of the particle
by a sphere with a radius $r_0,$ where $r_0$ agrees with the hadron's size. Since,
the external field (2.1) is not depend on $r,$ we have got the same equation for $R(r)$ as for a freely moving particle enclosed in a sphere, differing only from the expression by $%
k^2$ constant. From these boundary conditions one finds quantized energy
levels of spectrum branchs 
\begin{equation}
\label{2.13} 
\begin{array}{c}
\left( E_l^{(N)}\right) _{1,2}^2=\left( 
\frac{\alpha _l^{(N)}}{r_0}\mp \frac{g\tau ^{\frac 12}}2\right) ^2+M^2, \\ 
\left( E_l^{(N)}\right) _{3,4}^2=\left( \sqrt{\left( \frac{\alpha _l^{(N)}}{%
r_0}\right) ^2+g^2\tau }\mp \frac{g\tau ^{\frac 12}}2\right) ^2+M^2. 
\end{array}
\end{equation}
In this case the boundary condition on $R_l^\nu (r)$ with fixed value of the angular
momentum quantum number $l$ chooses from the continous energy spectrum (2.6) the
discrete series (2.13) of the radial quantum number $N$ for this value of $l$,
and the energy degeneracy for this state remains infinite. The
replacement connecting discrete and continous spectra is $\frac{\alpha
_l^{(N)}}{r_0}=\mid \overrightarrow{p}\mid $. By means of spectra (2.13) one
finds the same values for $k_\nu $ 
$$
k_\nu =k_l^{\left( N\right) }=\frac{\alpha _l^{(N)}}{r_0}. 
$$
The radius $a$ of the particle could be found from (2.10) using the maximum
condition $R^{\prime }(a)=0$ and is equal to%
$$
a_l^{(N)}=r_0\frac{\sqrt{l\left( l+1\right) }}{\alpha _l^{(N)}}. 
$$
As the energy spectrum is described by (2.13), the motion orbits of the particle are quantized
and defined by the orbital and radial quantum numbers $l$ and $N$ as well. Thus,
general solution of (2.7) is 
\begin{equation}
\label{2.14}\psi _j^{(i)}=\sum_{l=0}^\infty \sum_{m=-l}^lC_l^mY_l^m\left(
\theta ,\varphi \right) \frac{r_0}{\alpha _l^{(N)}\sqrt{r}}J_{l+1/2}\left(
kr\right) . 
\end{equation}
In the case of this field configuration the particle moves on ''$s","p","d","f"....$
orbitals known from orbital angular momentum eigenfuctions. For the continous
spectrum the constant $k$ and radius $a$ are%
$$
k=\left| \overrightarrow{p}\right| ,\ a_l\left( p\right) =\frac{\sqrt{l(l+1)}
}{\left| \overrightarrow{p}\right| }. 
$$
The plane wave could be decomposed by means of spherical waves and therefore, plane
wave solution obey initial equation as well [13].

Using the energy spectra (1.16) and (2.13), estimates for field strength of gluon condensate [2] and hadron's size the energy of a photon emitted by excited hadron may be calculated.
Using the wave functions (1.17) and (2.14) the explicit
expressions of Green's functions for these particles in finite temperature can be found 
[10], which gives a possibility to investigate different radiation corrections
in the chromomagnetic field and the quark condensate 
\begin{equation}
\label{2.15}<\overline{\psi }\psi >=-i\lim _{z\rightarrow 0}Tr\left[ G\left(
z\right) -G\left( z\right) \mid _{A=0}\right]. 
\end{equation}
The spectra found, (1.16) and (2.13) and the quark condensate (2.15) open new
possibilties to investigate chromomagnetic catalysis of color
superconductivity as well [6].

\bigskip

\renewcommand{\theequation}{A.\arabic{equation}} \hspace{-8mm} {\large {\bf %
Appendix}} \setcounter{equation}{0}\\ \newline

It is easy to reduce (1.10) to the form (1.12) in momentum
representation 
$$
\psi _{\pm }^{(i)}\left( \overrightarrow{r}\right) =\int_{-\infty }^{+\infty
}\frac{d^3p}{\left( 2\pi \right) ^{\frac 32}}e^{i\overrightarrow{p} 
\overrightarrow{r}}\widetilde{\psi }_{\pm }^{(i)}\left( \overrightarrow{p}%
\right) . 
$$
Then (1.10) will have the form 
\begin{equation}
\label{A.1}\left[ \left( \overrightarrow{p}^2+\frac{g^2\tau }%
2+M^2-E^2\right) ^2-g^2\tau \left( p_{\bot }^2+\frac{g^2\tau }4\right)
\right] \widetilde{\psi }_{\pm }^{(i)}\left( \overrightarrow{p}\right) =0. 
\end{equation}
which is divided into two equations%
$$
\left[ \left( \overrightarrow{p}^2+\frac{g^2\tau }2+M^2-E^2\right) \pm g\tau
^{1/2}\sqrt{p_{\bot }^2+\frac{g^2\tau }4}\right] \widetilde{\psi }_{\pm
}^{(i)}\left( \overrightarrow{p}\right) =0 
$$
and can be written in the following form
\begin{equation}
\label{A.2}\left( \sqrt{p_{\bot }^2+\frac{g^2\tau }4}\pm \frac{g\tau ^{\frac
12}}2\right) ^2\widetilde{\psi }_{\pm }^{(i)}\left( \overrightarrow{p}%
\right) =\left( E^2-M^2-p_3^2\right) \widetilde{\psi }_{\pm }^{(i)}\left( 
\overrightarrow{p}\right) . 
\end{equation}
From (A.2) we get 
$$
\sqrt{p_{\bot }^2+\frac{g^2\tau }4}\left( \widetilde{\psi }_{\pm
}^{(i)}\left( \overrightarrow{p}\right) \right) ^{1/2}=\left( \sqrt{%
E^2-M^2-p_3^2}\mp \frac{g\tau ^{\frac 12}}2\right) \left( \widetilde{\psi }%
_{\pm }^{(i)}\left( \overrightarrow{p}\right) \right) ^{1/2} 
$$
Squaring the last equality and returning to the coordinate representation by
an inverse Fourier transformation we get equation (1.12)%
$$
\left( -\nabla _{r,\varphi }^2+\frac{g^2\tau }4\right) \psi _{\pm
}^{(i)}=\left( \sqrt{E^2-M^2-p_3^2}\mp \frac{g\tau ^{\frac 12}}2\right)
^2\psi _{\pm }^{(i)}. 
$$

\begin{center}
{\large {\bf Acknowledgements}}
\end{center}

The author thanks IPM for financing his visit to this institute and Prof. F.Ardalan for his hospitality during this visit.

\end{document}